\documentclass[aps,epsfig,preprint]{revtex4-1}
\usepackage{soul}
\usepackage[section]{placeins}        
\usepackage{amsmath}
\usepackage{amssymb}
\usepackage{graphics}
\usepackage{caption}
\usepackage{wrapfig}
\usepackage{sidecap}
\usepackage{epsfig}
\usepackage{pdfpages}
\usepackage{float}
\usepackage{subcaption}

\newcommand{\be}{\begin{equation}}
\newcommand{\ee}{\end{equation}}

\newcommand{\bea}{\begin{eqnarray}}
\newcommand{\eea}{\end{eqnarray}}

\begin{document}
\title{Visco-elastic fluid simulations of coherent structures in strongly coupled dusty plasma medium }
\author{Vikram Singh Dharodi}
\author{Sanat Kumar Tiwari}
\author{Amita Das}
\email{amita@ipr.res.in}
\affiliation{Institute for Plasma Research, Bhat , Gandhinagar - 382428, India }
\date{\today}
\begin{abstract}
A generalized hydrodynamic  (GHD) model depicting the behaviour of visco-elastic fluids
has often been invoked to explore the behaviour of a strongly coupled dusty plasma medium
below their crystallization limit. The model has been successful in describing the
collective normal modes of the strongly coupled dusty plasma medium observed experimentally.
The paper focuses on  the study of nonlinear dynamical characteristic features  of this model.
Specifically, the  evolution of coherent vorticity patches are being investigated here within the
framework of this  model. A comparison with  Newtonian
fluids and Molecular Dynamics (MD) simulations treating the dust species interacting through the
 Yukawa potential has also been presented.
\end{abstract}
\maketitle
\section{Introduction}

The understanding of a strongly coupled state of matter is a  problem of frontier research in
physics \cite{fortov_book,bannasch_13,horn_91}. While such a state of matter can be found only
under extreme conditions of temperature and densities, the dusty plasma proves to be a unique
system where the strong coupling conditions can be achieved easily even at normal pressure and
temperature due to the high charge it possesses. This has attracted significant research interest
in the field of dusty plasma medium \cite{Merlino2004,Thomas}.

A phenomenological visco-elastic fluid model, known as the Generalized Hydrodynamic (GHD) model,
has often been invoked for the study of the dusty plasma medium in the intermediate coupling regime
of $1< \Gamma < 170$, (here $\Gamma$ is the ratio of the inter-particle potential energy to the
kinetic energy) \cite{frenkel_kinetic,Kawghd98}.
In this regime the dust species do not crystallize and neither do they behave like normal Newtonian
fluids. This model has been successful in  predicting the dispersion characteristics of the
transverse shear (TS) wave in the medium \cite{Kawghd98,Kaw2001}, which have been experimentally
demonstrated \cite{PhysRevLett.88.175001}. The mode dispersion has also been obtained within
the Molecular Dynamics (MD) simulations which treat the dust particles as interacting through
Yukawa potential \cite{PhysRevLett.84.6026}.

The purpose of the paper is to investigate the nonlinear regime of the GHD model. In particular
we intend to study the  evolution and interaction of coherent structures which have an important
role to play  in defining transport properties of any medium.
The elasticity effects in GHD  is introduced  through  a memory relaxation parameter
$\tau_m$ \cite{Kawghd98, berkovsky_92, Ichimaru87} in the  evolution equation for velocity.

As the governing dynamics of strongly coupled dusty plasma medium is different
than that of Newtonian fluids, the existence of various coherent structures as
well as their stability and evolution in such medium may have significant
differences. We have chosen an important phenomena of vortex evolution and
addressed its application to strongly coupled dusty plasma medium.
In past the phenomena of vortex evolution has been studied in a range of
 physical systems like hydrodynamic fluids \cite{Weiss_turbulence},
 planetary atmospheres and their convective interiors,
 electron plasma \cite{PhysRevLett.67.588,Driscoll200221}, etc.
The dust vortices have also been studied in presence as well as in absence of
magnetic field \cite{Kaw_rotation_2002,Shukla,Law_1998,Morfill_1999,Klindworth_2000,Morfill_2004}.
Recently Yoshifumi et al. have presented interesting experimental results on
dust rotation \cite{yoshifumi_13}.
Konopka et al. and Sato et al. have shown the rotation of dust particles experimentally in the
magnetized dusty plasma \cite{Konopka_2000,Sato_2001}.
Schwabe et al. reported in their dusty plasma experiment, formation of variety of rotating dust structures
depending on varying magnetic field strength \cite{Schwabe_2011} and further
numerically studied the vortex movements by adding some
micro particles around the void in complex plasma simulation \cite{schwabe_prl14}.
The coherent solutions in the form of  tripolar vortex  have also been studied theoretically
in the context of dusty plasma \cite{Vranješ1999317}.

We present the 2-D numerical simulation studies  of  the GHD equations  to study
the evolution of various configurations of vorticity patches. It is observed that
in contrast  to Newtonian fluids there is an  emission of transverse shear waves
from the vorticity patches due to which the strength of the coherent structure gradually
 diminishes. An interesting  observation is that the interplay between the  emitted
 transverse shear wave  and the coherent vorticity structures ultimately results in
 a proper mixing of the fluid.

 The manuscript has been organized as follows. Section II contains the details of
 the governing equations. In section III a brief description of the  numerical approach
 and some simplified cases of  simulations with expected results  have been discussed
 which validate the numerical  code.  In section IV we present the evolution of
various  different configuration of vorticity patches. In particular the role of
transverse shear wave in the evolution is clearly demonstrated. For certain sharp
vorticity profiles the interplay of the Kelvin Helmholtz instability and the transverse
shear wave has also been shown. The comparison with the evolution of Newtonian fluids
show that in all cases the GHD fluid shows a better mixing. In section V a comparison
is provided with the Molecular Dynamics (MD) simulation results.
Finally, section VI contains the discussion and conclusion.

\section{Governing Equations}
A typical plasma which is a   collection of electrons and ions, is generally found in
a state of weak coupling wherein $\Gamma$ the coupling parameter defining the ratio of
 average inter-particle potential  energy  to the  average thermal energy associated
 with the particles, is less than unity.
  However, when micron sized  dust particles are sprinkled in such a plasma, it
  typically acquires a very high charge due to constant impingement and attachment of
  the lighter electron and ion species. In general the lighter electron flux on the
  dust grain surface is comparatively higher than that of the  ion flux,  causing the
  dust species to acquire a net negative charge.
  In such a system the dust component acts as a third species with a very high
  negative charge. Due to the large charge on the dust particles,
  this species can often be  easily found in the strong coupling regime corresponding
  to the coupling parameter $\Gamma >> 1$. Ikezi {\it et al.} \cite{ikezi_86} have
  shown  that when the coupling parameter ${\Gamma \gtrsim  170}$, the dust particles
  get organized in regular lattice points forming a Coulomb crystal. However, for
  $1< {\Gamma < 170}$ the dusty plasma system has a behaviour which is intermediate
   to that of fluid and solids and has in the past been looked upon as a visco-elastic
   system. It is this regime of dusty plasma system that we would be exploring here.
   The visco-elastic response is typically modeled by a generalized hydrodynamic description
  in which the elasticity is represented by a memory relaxation parameter $\tau_m$ \cite{frenkel_kinetic}.
  Generalized hydrodynamic (GHD) fluid model provides a description of dust fluid in both weak
(simple charged fluid) and strong coupling limits (visco-elastic fluids). These two aspects
are combined in terms of a characteristic time scale {$\tau_m$} i.e. memory
relaxation time.

The visco-elastic description of the electrostatic response of strongly coupled dusty plasma
medium is provided by the following coupled set of continuity equation, the evolution of
velocity through a Generalized Hydrodynamic description and the Poisson's equation respectively.
{\small
\begin{equation}\label{eq:continuity}
  \frac{\partial n}{\partial t} + \nabla \cdot \left(n\vec{v}\right)= 0
  \end{equation}
	 }
  {
\begin{equation}\label{eq:momentum}
 \left[1 + \tau_m {\frac{d}{dt}}\right]\left[ {{n}\frac{d\vec{v}}{dt}}+{\nabla P}-{n} \nabla \phi \right]
 = \eta \nabla^2 \vec{v}+{\left({\zeta+\frac{\eta}{3}}\right)}{\nabla}{\left(\nabla \cdot \vec{v}\right)}
  \end{equation}
	 }%
 {\small
 \begin{equation}
 \nabla^2 \phi  = \left[n + \mu_{e}exp(\sigma_e\phi) - \mu_{i}exp(-\phi)\right]
\end{equation}
	}%
%
Here, $\vec{v}, n$ and $\phi$ are the normalised dust fluid velocity, dust density and
potential respectively in dusty plasma medium.
The total time derivative is represented as $\frac{d}{dt}= \left(\frac{\partial} {\partial t}+\vec{v}  \cdot \nabla \right)$.
The above equations has been presented in normalised form \cite{Sanat-KH-inst-2012}
with parameters $\sigma_e = \frac{T_i}{T_e}$, $\mu_{e} = \frac{n_{e0}}{Z_{d}n_{d0}}$
and $\mu_{i} = \frac{n_{i0}}{Z_{d}n_{d0}}$.
Dust particles carrying a negative charge $Z_d$ and $n_{s0}$ ($s=e,i,d$) denotes the
normalizations of the number density of the specific species and the equilibrium
quantities are  denoted by the subscript $0$. The memory effect related to elasticity is
incorporated through a relaxation time parameter ${\tau_m}$,
and $\eta $, $ \zeta $ are the shear and bulk viscosity coefficients respectively.
If {${\tau_m}{\frac{\partial}{\partial{t}}}< 1$}, there are no memory effects
and the equation of motion is that for a standard viscous fluid driven by self consistent
electric and pressure fields	and for { ${\tau_m}{\frac{\partial}{\partial{t}}} \geq 1$}
, memory effects are strong (for time scales of interest, each fluid element remembers where
it came from) and the viscosity coefficient $\eta$ becomes more like a non dissipative
elastic coefficient supported by strong particle correlation.
The inertia of electrons and ions is negligible  at slow dust time scales
and hence these species can be assumed to follow a Boltzmann distribution.
\section{Numerical implementation and validation}
In this manuscript  we  concentrate on studying the in-compressible limit of the above
set of equations i.e. we assume that $\nabla \cdot \vec{v} = 0$.
The coupled set of governing eqs. (\ref{eq:continuity})-(\ref{eq:momentum})  have been recast in the
following form in this limit.

\begin{eqnarray}\label{eq:vorticity}
\frac{\partial{\vec \xi}} {\partial t}+\left(\vec{v} \cdot \vec \nabla\right){{\vec \xi}}={\vec \nabla}{\times}{\vec \psi}
\end{eqnarray}
\begin{eqnarray}\label{eq:psi}
\frac{\partial {\vec \psi}} {\partial t}+\left(\vec{v} \cdot \vec \nabla\right){\vec \psi}=
{\frac{\eta}{\tau_m}}{\nabla^2}{\vec{v} }-{\frac{\vec \psi}{\tau_m}}
\end{eqnarray}
Here $\eta $ is now termed as the kinematic viscosity, $ {\vec \xi}$ is the vorticity,
given as $ {\vec \xi}={\vec \nabla}{\times}{\vec{v}} $
and the velocity at each time step is updated by using the poisson's equation $ {\nabla^2}{\vec{v}}=
-{\vec \nabla}{\times}{\vec \xi}$.
Also, ${\vec \psi}={\frac{d\vec{v}}{dt}+\frac{\nabla P}{n}-\nabla \phi}$
and its curl gives eq.~\ref{eq:vorticity}.
It should be noted that in this particular limit there is nothing specific which is suggestive of
the fact that the system corresponds to a strongly coupled dusty plasma medium.
Thus,  the results presented in this manuscript would in general be applicable to any visco-elastic
medium and not be restricted to the strongly coupled dusty plasma medium.

The linearization of the above set of equation yields the
following dispersion relation for the  transverse shear waves \cite{Kawghd98},
{
\begin{equation}
\omega = \frac{-{i}{\eta}{k^2}}{1-{i}{\omega}{\tau_m}}
\end{equation}
	}
In the strong coupling limit of   ({$\omega \tau_m >>1$}) this yields
\begin{equation}
{\frac{\omega}{k} = \sqrt{\frac{\eta^{}}{\tau_m}}}
\label{trans_disp}
\end{equation}
which implies wave propagation and in the other limit
of  ({$\omega \tau_m << 1$})
\begin{equation}
\omega = -{i}{\eta}{k^2} \nonumber
\label{trans_disp}
\end{equation}
we have the usual damping  due to viscosity in  hydrodynamic fluids.


We have used the flux corrected scheme (Boris {\it et al.} \cite{boris_book})
to evolve the coupled set of eqs. (\ref{eq:vorticity})-(\ref{eq:psi}).
The equations were evolved for a slab sinusoidal perturbation and the dispersion
relation for the transverse shear wave was verified numerically  as a part of
code validation \cite{Sanat-AIP-sheared-flow}.

\section{Dynamical evolution of  vorticity patches through GHD model}
A typical fluid flow contains a wide  variety of coherent patterns in
the form of localized vorticity  patches. Their interaction and evolution are important
for the understanding of the system which  in turn is responsible for the transport
properties
of the system. The objective of the present work is to understand  the dynamical
characteristics of these entities for a strongly coupled system  within the framework
of the visco-elastic GHD model. 

We consider the following specific cases in particular: (i) evolution of circular and
elliptical vorticity patches and (ii) Interaction amidst vorticity patches of like and
unlike signs. The vorticity patch representing a sheared rotation emits the transverse
shear wave for a visco-elastic fluid, making the evolution  dramatic in terms of
rapid mixing and transport behaviour of the GHD fluid system.

\subsection{Evolution of vorticity patches}
We consider two cases of  circularly rotating fluid profile, namely
 having  (A)  smooth rotational profile and consider another which
 has a  (B) sharp cut off.
In the former case (A) the vorticity smooth profile (rotating in a clockwise direction) is given by,
${\xi_{0}(x,y,t_0)}={\Omega_0}exp\left(-\left({\left(x-x_c \right)^2+(y-y_c)^2}\right)/{a^2_c}\right)$.
Here $ {\Omega_0}={\Gamma_0}/{\pi{a^2_c}}$, $ {\Gamma_0} $ is the total circulation,
$a_c$ is the vortex core radius and the numerical simulation has been carried
out for ${a_c}$=1.5, ${\Omega_0}=8$ and ${x_c}={y_c}$=0.

\begin{figure}[!htb]
                \includegraphics[width=\textwidth]{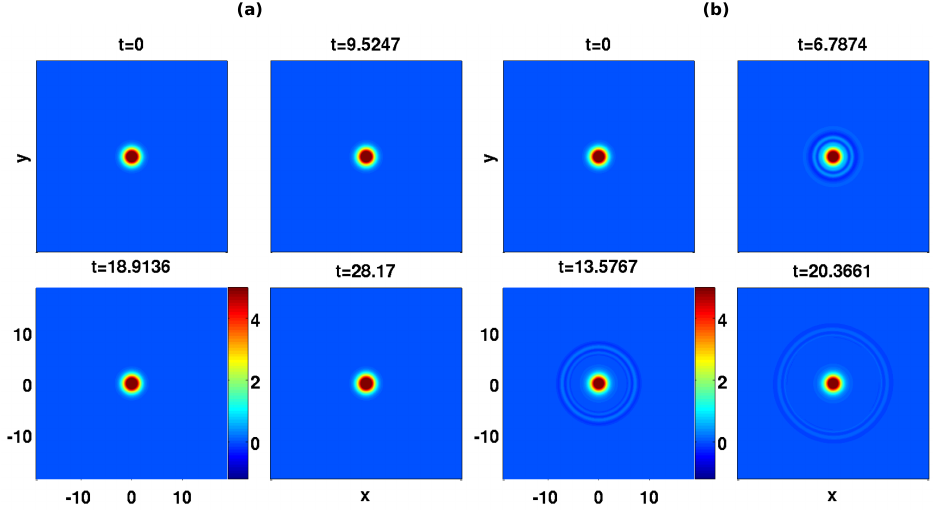}
        \caption{Evolution of smooth circular vorticity profile in time for (a) hydrodynamic fluid and (b)
        visco-elastic fluid with parameters ${\eta}=5$, ${\tau_m}=20$.}
          \label{fig:smooth_vorticity}
\end{figure}

\begin{figure}
         \centering
        \includegraphics[width=\textwidth]{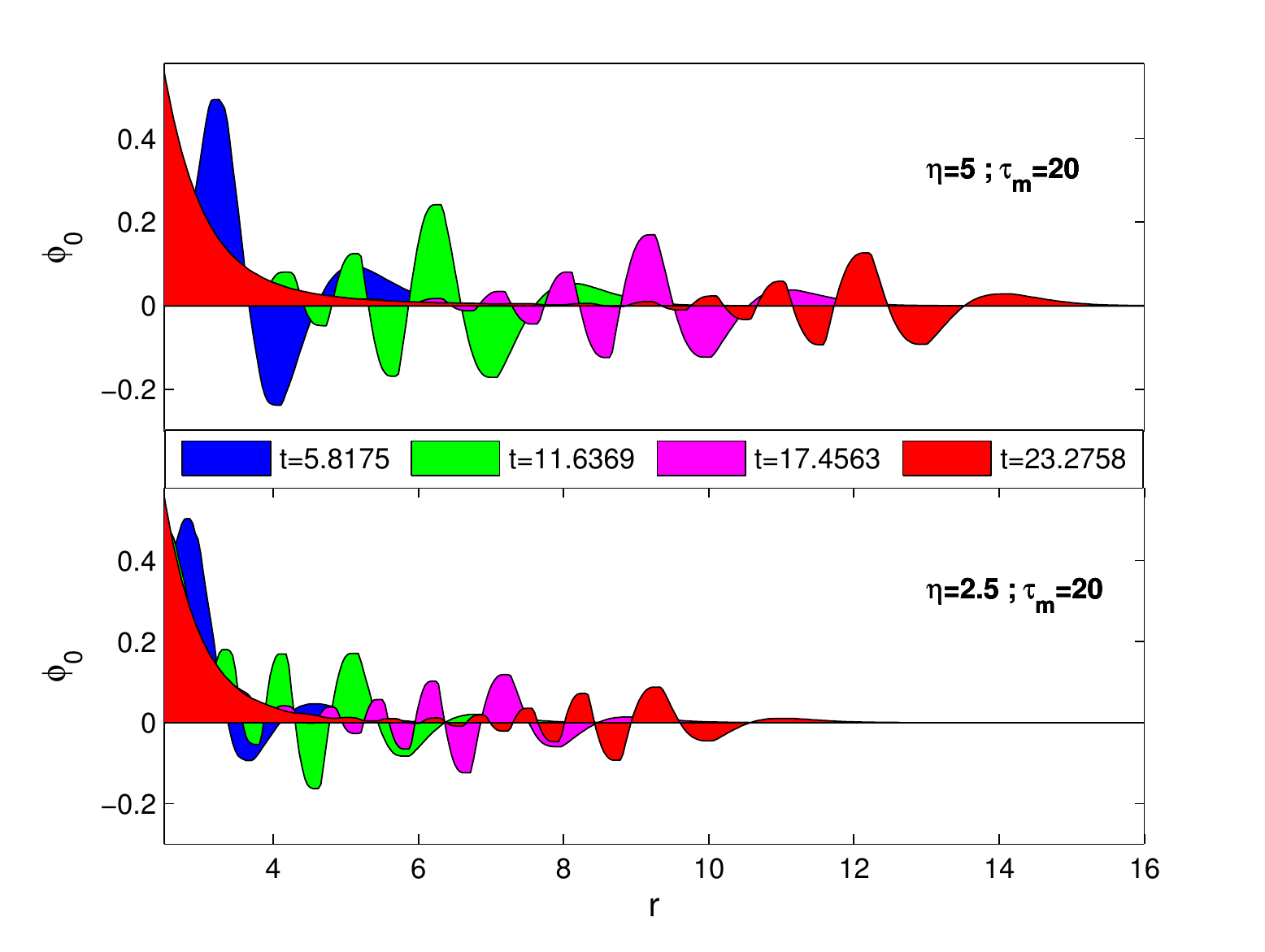}
      \caption{Radial emission (emerging wavefornt) of TS waves along one of the axes at
      different times during vortex evolution in visco-elastic medium for the parameters
      (a) ${\eta}=5, \tau_m=20$ and (b) $ {\eta}=2.5 $, ${\tau_m}=20$.}
        \label{fig:radial_emission}
\end{figure}

\begin{figure}[!h]
        \centering
                \includegraphics[width=\textwidth]{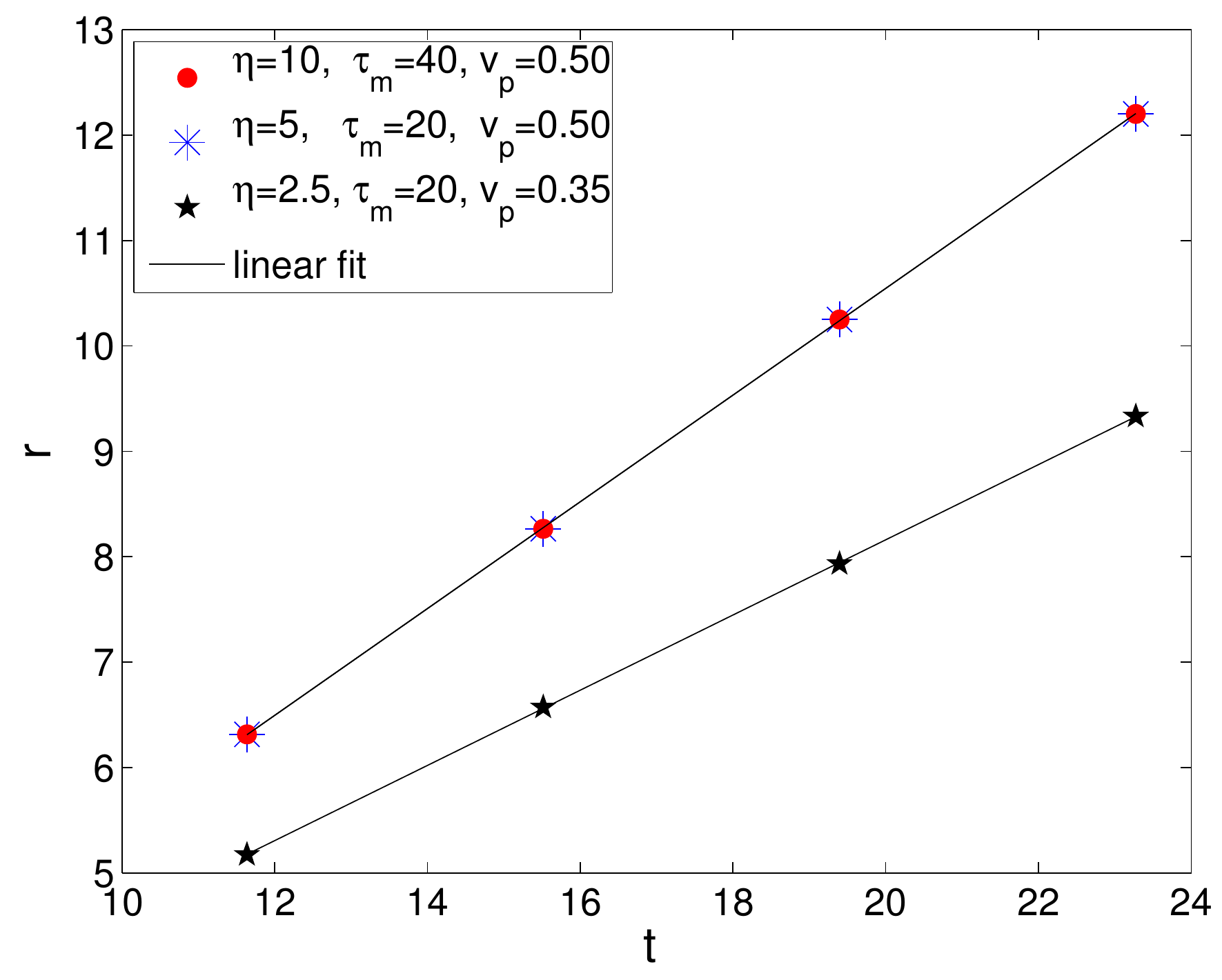}
                \caption{ Wavefront position of TS waves at different time steps with
                          parameter values  $ {\eta}=10, {\tau_m}=40$ {$ (\color{red} \bullet  $)},
                          $ {\eta}=5, {\tau_m}=20$ ({$\color{blue}{\ast}$)} and
                          $ {\eta}=2.5, {\tau_m}=20$ {$ (\color{black} \bigstar$)}
                          , where $v_p$ is the phase velocity
                           related to corresponding parameters (of corresponding color) and black line is
                           linear fitted curve.}
          \label{fig:transverse_wave}
 \end{figure}

\begin{figure}[!h]
                \centering
                \includegraphics[width=\textwidth]{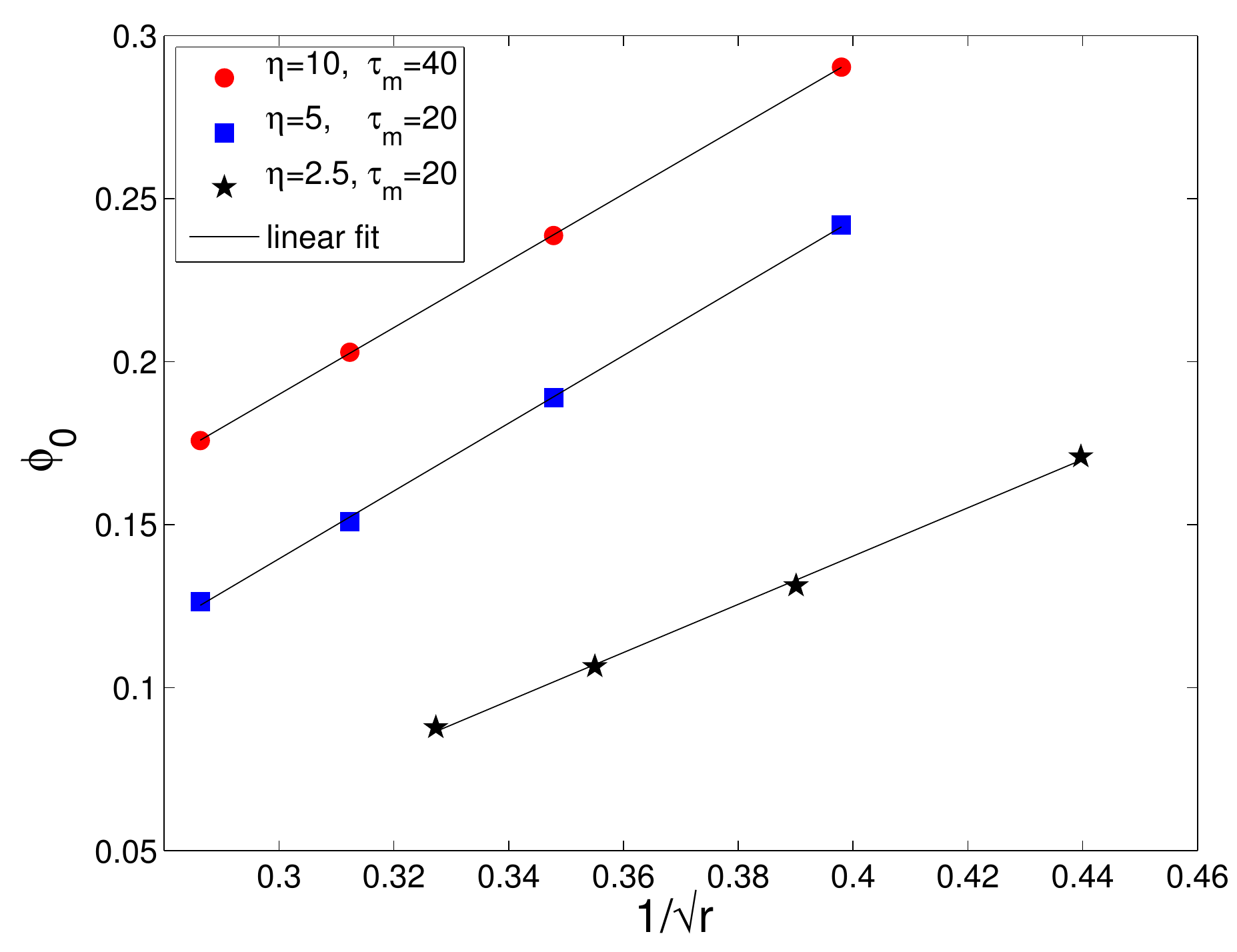}
               \caption{ Wavefront amplitude of TS waves with {${1}/{\surd r}$} with
                          parameter values $ {\eta}=10, {\tau_m}=40$ {$ (\color{red} \bullet$)} $,
                          {\eta}=5, {\tau_m}=20$ ({$\color{blue}{ \blacksquare} $)}
                          and ${\eta}=2.5, {\tau_m}=20$ {$(\color{black} \bigstar$)}
                           with black line as linear fitted curve.}
          \label{fig:cylindrical_wave}
\end{figure}

  In fig.~\ref{fig:smooth_vorticity} we show  a comparison of  the evolution of such a circular
  vorticity patch for a hydrodynamic fluid and the GHD system, through the color contour plot of
  vorticity.  It can be observed that while the structure is stable in the context of the
  hydrodynamic system, for the GHD case there is a radial emission of waves.
  The radial emission can be seen more clearly in fig.~\ref{fig:radial_emission}
  where the vorticity profile as a function of one of the axis (say $x$) has been
  plotted for various times with different line styles. A perturbation clearly
  proceeds outwards. The radial speed of this perturbation has been evaluated and found
  to match with $\sqrt{\eta/\tau_m}$ as has been shown in fig.~\ref{fig:transverse_wave}.
  The circular nature of the emitted wave also suggests that the amplitude of these
  characteristic perturbations  should display a $1/\sqrt{r}$  radial fall off.
  This has also been demonstrated numerically as shown in the plot of fig.~\ref{fig:cylindrical_wave}.

\begin{figure}[!h]
        \centering
                \includegraphics[width=\textwidth]{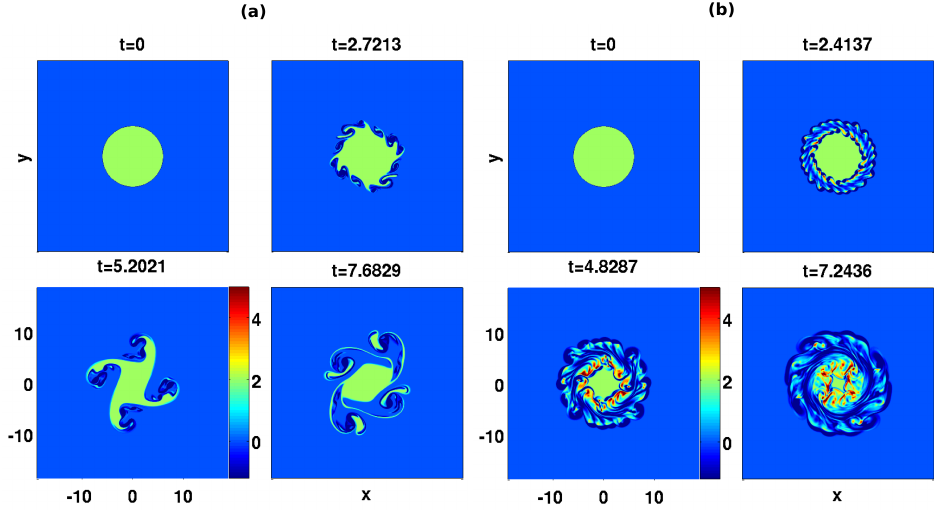}
        \caption{Evolution of circular sharp vorticity profile in time for
        (a) hydrodynamic fluid and (b) visco-elastic fluid with parameters ${\eta}=5, {\tau_m}=20$.}
                  \label{fig:sharp_hd_ghd}
\end{figure}

\begin{figure}[!h]
        \centering
                \includegraphics[width=\textwidth]{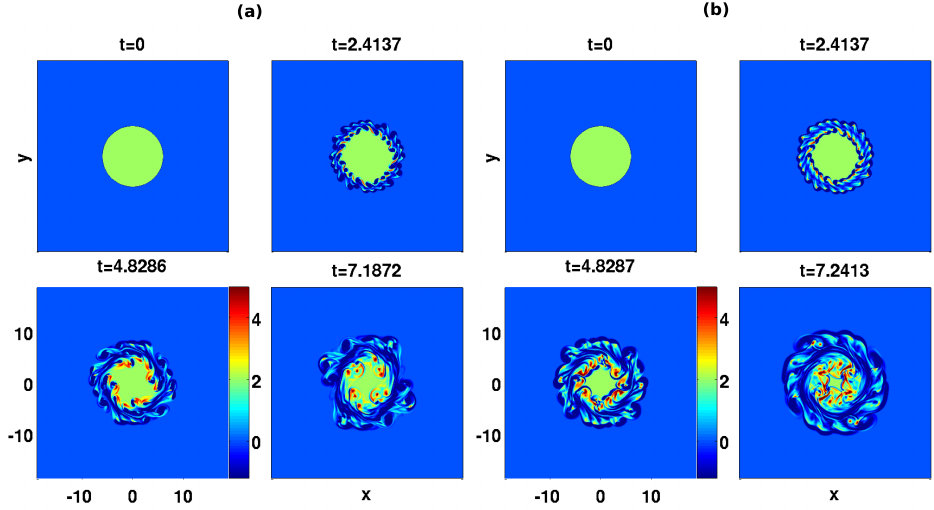}
       \caption{ Evolution of sharp circular vorticity profile in time for
			 strongly coupled dusty plasma medium for the cases
        (a) ${\eta}=2.5, {\tau_m}=20$, and (b) ${\eta}=10, {\tau_m}=40$.}
                          \label{fig:sharp_ghd_compare}
\end{figure}

We next choose profile B with a sharp cut off, i.e.
 setting the vorticity ${\xi}_{z0} = 0$ beyond $r = r_0$ ($ =6.0$).
 For $r \leq r_0$ the vorticity is  chosen to have a  constant value (${{\xi}_{z0}}=2$).
 The abruptness of the vorticity profile generates a
 strong rotational sheared flow, which in turn is drastically unstable to the Kelvin
 Helmholtz instability for both the hydrodynamic (HD) fluid as well as the GHD system.
 The vorticity contours at various times for both HD and GHD system
 have been shown in fig.~\ref{fig:sharp_hd_ghd}.
 It should be noted that perturbations in the context of HD
 system remain considerably  localized when compared to those of GHD. Basically the
 fluid is much less perturbed in the context of HD than that of GHD. This can be
 understood by noticing that for the GHD system the strong intermixing occurs due to
 the emission of the transverse shear waves. There is a clear propagation of two
 transverse shear wave fronts, one inward and the other outward and a concomitant KH
 destabilization at each of these fronts.  The radial shear wave is clearly
 instrumental in efficient mixing of the fluids entrained inside the vortex structure
 with the  outside fluid. For HD system the initial KH perturbation evolves towards a
 very anisotropic isolated structure. The mixing of the fluid in this case is clearly minimal.
 In fig.~\ref{fig:sharp_ghd_compare} we have compared two different cases of
 GHD simulation, with different values of $\eta$ and $\tau_m$ parameter
 ({ {${\eta}=2.5, {\tau_m}=20, v_p$=0.35} for fig.~\ref{fig:sharp_ghd_compare}(a) and
 {${\eta}=10, {\tau_m}=40, v_p$=0.5} for fig.~\ref{fig:sharp_ghd_compare}(b)}).
 We observe that the mixing is better for higher phase velocity of the TS wave.
Also, the figs.~\ref{fig:sharp_hd_ghd}(b),~\ref{fig:sharp_ghd_compare}(b)
show that the vortex evolution is similar in time for same value of TS wave
phase velocity ({$v_p={\sqrt{{\eta^{}}/{\tau_m}}}$}=0.5)

\begin{figure}[!h]
        \centering
                    \includegraphics[width=\textwidth]{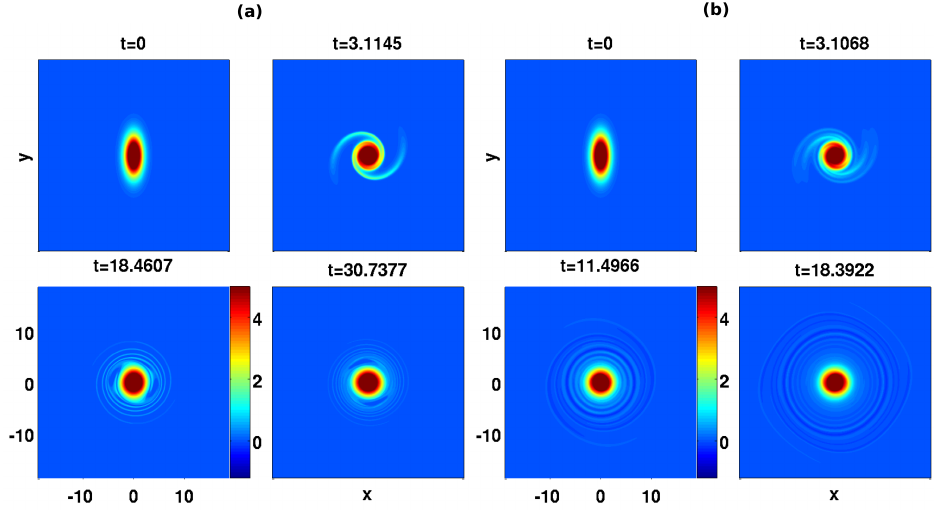}
       \caption{Evolution of elliptical vorticity  profile in time for
       (a) hydrodynamic fluid and (b) visco-elastic fluid with parameters $ {\eta}=5, {\tau_m}=20$.}
               \label{fig:elliptical_hd_ghd}
       \end{figure}

\begin{figure}[!h]
        \centering
                \includegraphics[width=\textwidth]{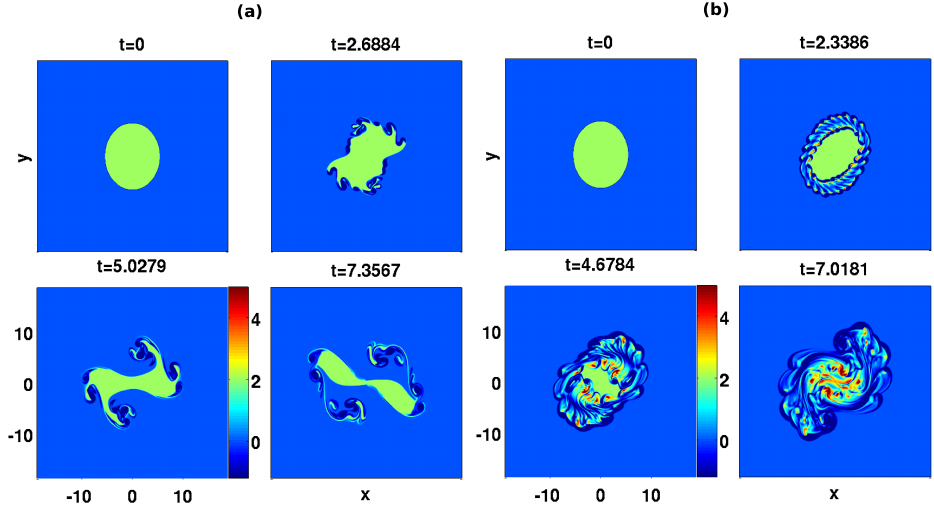}
        \caption{Vorticity evolution for sharp elliptical profile in time for
        (a) hydrodynamic fluid and (b) visco-elastic fluid with parameters $ {\eta}=5, {\tau_m}=20$.}
                  \label{fig:sharp_hd_ghd_ellipse}
\end{figure}

 Often the vorticity structure in a fluid may not have a circular shape. We consider,
 therefore, for our studies an initial  distorted patch of vorticity. A simple elliptical
 form of distortion have been considered by us. Various time frames of the evolution of
 such a vortex pattern in both HD and GHD case has been shown in
 figs.~\ref{fig:elliptical_hd_ghd},~\ref{fig:sharp_hd_ghd_ellipse}
 for smooth and sharp vorticity profiles respectively. It can be seen that while
 the distorted shape of the vorticity patch does help somewhat in making the
 transport better in the context of HD, the GHD case still proves to be more
 efficiently mixing the fluids.

 \subsection{Interaction amidst vorticity patches}
We have also investigated the process of interaction between various
 vortex structures within the GHD formalism for a strongly coupled medium.
 A correspondence with HD system has also been provided.

\begin{figure}[!h]
        \centering
                    \includegraphics[width=\textwidth]{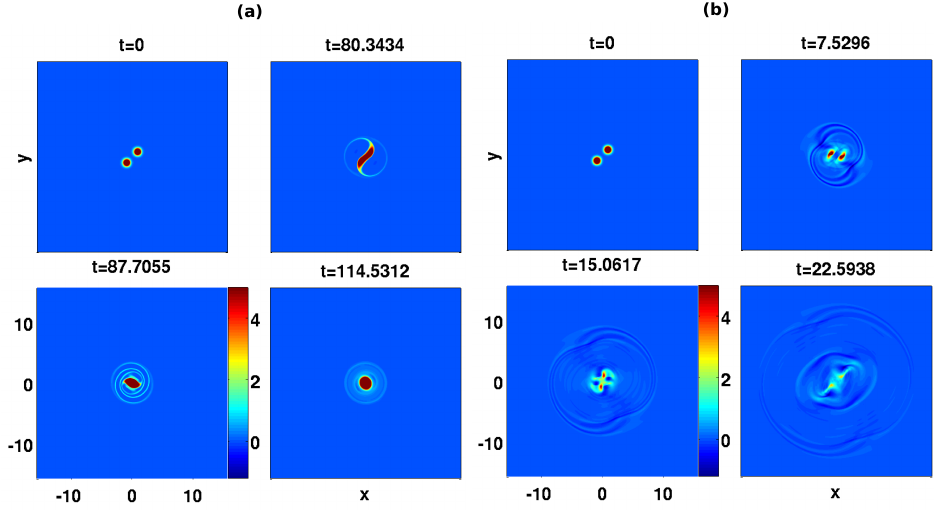}
         \caption{Evolution of two like sign vortices in time for
         (a) hydrodynamic fluid and (b) visco-elastic fluid with parameters $ {\eta}=5, {\tau_m}=20$.}
              \label{fig:merger_hd_ghd_0.7}
                    \end{figure}

The interaction and subsequent merging of two like signed vorticity patches
have been well known in the context of a hydrodynamic system
\cite{Meunier-merging,Meunier2005431,Josserand2007779}.
The same in the case of GHD
has been illustrated in the subplots of fig.~\ref{fig:merger_hd_ghd_0.7}.
In contrast to HD, the merging does not lead to a coherent final form,
instead as expected the TS waves continue to dominate the system.

\begin{figure}[!h]
        \centering
                    \includegraphics[width=\textwidth]{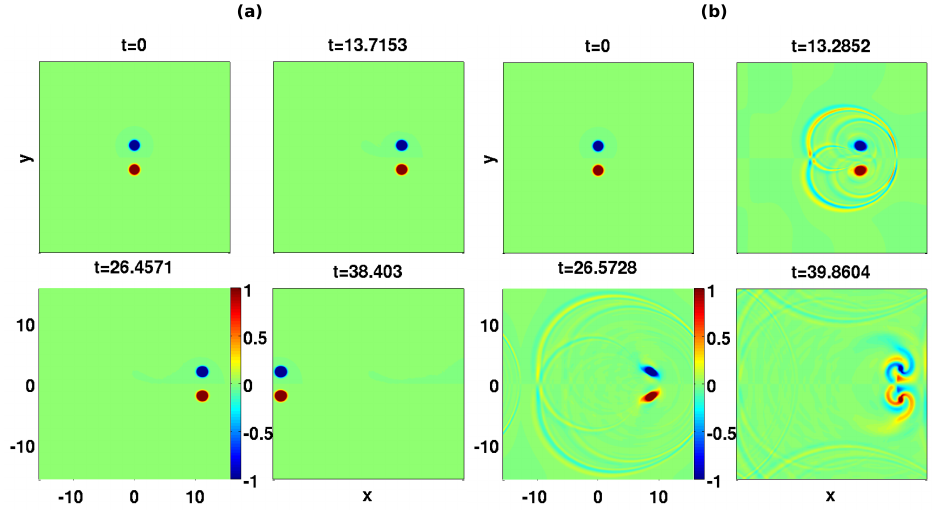}
         \caption{Evolution of two unlike sign vortices with time for
         (a) hydrodynamic fluid and (b) visco-elastic fluid with parameters $ {\eta}=5, {\tau_m}=20$.}
               \label{fig:dipole_hd_ghd}
       \end{figure}

The unlike vortices, when brought together in the context of HD system are observed
to propagate together along the direction of their axis as stable dipoles.
In GHD case too while the structures do propagate together axially
for some time as shown in fig.~\ref{fig:dipole_hd_ghd}.
The  continuous emission of waves, however, distorts the structure ultimately.

The emission of transverse shear waves appears to have a  predominant role in the
mixing and transport of the fluid elements in the context of the visco-elastic GHD
system. The strong mixing can be suppressed provided the TS waves have damped
characteristics in the medium.

In the next section we investigate the specific case of the strongly coupled dusty
plasma medium by employing the MD simulations.

\section{Molecular dynamics studies }
We have carried out a Molecular Dynamics simulation of the dust particles in
the strongly coupled regime using the open source LAMMPS code
\cite{Plimpton1995}.
The role of electrons and ions has only been assumed to provide shielding effect to dust species causing effective
interdust interaction through a Yukawa inter particle dust potential.
\begin{equation}
 U(r) = \frac{Q^2}{4\pi\epsilon_0 r}\exp(-r/\lambda_D)
 \label{yukm}
\end{equation}
The parameters $Q$ and $\lambda_D$ are the charge on dust particle and plasma Debye
length respectively. We have performed 2-D simulations and periodic boundary conditions
have been employed for present simulations of vortex patches.
We have taken typical inter-dust distance $a_d = 0.418 \times 10^{-3}$m, the fixed
charge on dust particle is taken $Q = 11940e^{-}$, system size $l_x = l_y = 0.1672$m and dust
mass $m_d = 6.99 \times 10^{-13}$Kg.
As we have chosen $\kappa = a_d/\lambda_D =0.5$, the corresponding value of $\lambda_D = 8.36 \times 10^{-4}$m.
With these parameters, the dust plasma frequency comes out to be typically $\sim 35$Hz
corresponding to dust plasma period of $0.028$s. We have chosen the typical time step
$0.002$s which can well resolve the characteristic dust plasma frequency.
The parameters chosen for simulation are very relevant to dusty plasma experiments \cite{Nosenko2004}.
We equilibrated our system first as a NVT ensemble for 40s and then further as NVE ensemble for next 40s.
After this equilibration our system becomes ready for study at some particular coupling parameter $\Gamma$.
The expression for coupling parameter is $\Gamma = Q^2/\left(4\pi \epsilon_0 a_d k_B T_d\right)$.
For some fix values of charge $Q$ and dust density (hence typical interdust separation $a_d$), the dust particle's temperature
decides the value of coupling parameter. We have choosen some fix value of $\Gamma$ and then calculated the dust temperature corresponding
to it in our system. To achieve the system at this temperature, we have equilibrated it as NVT ensamble as described earlier.
For present simulations, we have not considered the effect of gas friction in our system as our primary aim is to look into
the effect of strong coupling over vortex dynamics. The effects due to gas friction is important factor and will be added in further
simulation studies to make results more closer to experiments.

\begin{figure}[!ht]
\centering
\includegraphics[height=10.0cm,width=12.0cm]{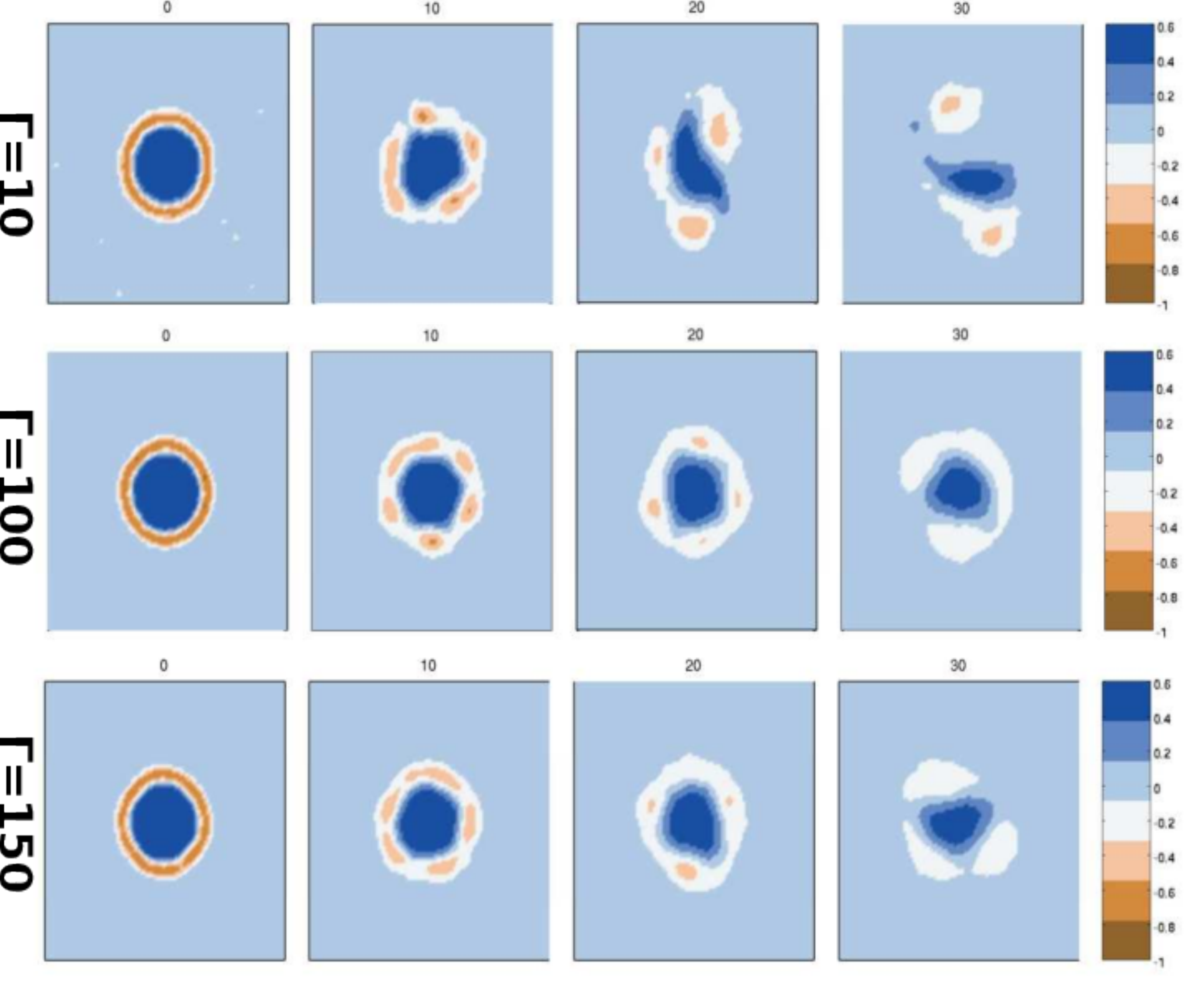}
\caption{Evolution of sharp vortex in strongly coupled dusty plasma
	   medium (Yukawa medium) at different coupling parameters
	    $\Gamma =10,100$ and $\Gamma=150$ and $\kappa = 0.5$.}
\label{vortex_md}
\end{figure}

All physical studies made on this system have been performed in NVE ensemble.
In our present studies, we have given a vortex like initial profile to a specific circular region of Yukawa
system making a sharp interface of vorticity between the circular patch region and the rest of the system.
The initial flows in circular patch have been chosen as $v_{x0} = -y$ and $v_{y0} = x$.
Figure~\ref{vortex_md} shows the evolution of sharp vortex patch for different values of coupling parameter $\Gamma$.
The formation of KH instability at the interface of vortex patch can easily be observed in all the cases.
But as the value of coupling parameter is increased, we found that the vorticity patches seem more
stable as the growth of KH instability reduces with increasing $\Gamma$.
This is the reason why the further evolution of a single vortex forms a tripolar structure for
small coupling parameter while similar structure remains reasonably stable for higher value of $\Gamma$.

\section{Summary and conclusion}
The evolution and interaction of localized vortex patterns for a strongly coupled
medium depicted by the visco-elastic GHD description have been studied.
Preliminary studies with the Molecular Dynamics simulations have also been carried out.

We observe that the  rotational sheared flow in a localized vortex patterns is susceptible
to the Kelvin - Helmholtz (KH) destabilization which is similar to the Newtonian fluids.
It is, however,  necessary that for  KH destabilization the shear should be strong
and have an inflection point.
This is possible when we considered the sharp cut off in the vorticity patches.
In contrast to the Newtonian fluid the GHD visco-elastic medium, in addition to KH also
permits the emission of radially (inward as well as outward) propagating transverse shear waves.
  The phase speed of the waves and the  $1/r$ fall in their intensity has  been verified numerically.

Our studies show that due to the existence of such transverse shear waves in the strongly
coupled medium  the mixing and transport behaviour in these fluids is much better
than Newtonian hydrodynamic systems.
The chances of fluid element and/or test particles to remain entrained for long duration
within a localized region is insignificant in GHD when compared to the Newtonian fluid.

We are in the process of quantifying this transport behaviour by carrying out test
particle simulations in the system of GHD model.
The numerical dispersion of these particles in a GHD flow would provide an estimate of
the diffusion in such a medium.
An analysis of the separation of the particle trajectories is also being  carried out
to understand the similarity and differences with the 2-D hydrodynamic system studied
by Falkovich et al \cite{falkovich_01}.
These observations would be presented in a subsequent publication.
\newpage
\bibliographystyle{unsrt}
\bibliography{vikram_vort}
\newpage
%
~~~~~~~~~~~~~~~~~~~~~~~~
\FloatBarrier
       \FloatBarrier
       \FloatBarrier
        \FloatBarrier
       \FloatBarrier
       \FloatBarrier
\end{document}